\documentclass[prd,aps,floats,twocolumn,nofootinbib]{revtex4-1}
\usepackage{slashed}
\usepackage{mathtools}
\usepackage{amsfonts}
\usepackage{amssymb}
\usepackage{epsfig}
\usepackage{empheq}
\usepackage{enumitem}
\usepackage{amsmath} 
\usepackage{blkarray} 
\usepackage{multirow}
\usepackage{graphicx} 
\usepackage{float} 

\usepackage{bm}

\usepackage[polutonikogreek,english]{babel}
\usepackage[or]{teubner}

\usepackage{xcolor}

\begin{document}
\newcommand*\widefbox[1]{\fbox{\hspace{2em}#1\hspace{2em}}}
\newcommand{\m}[1]{\mathcal{#1}}
\newcommand{\nn}{\nonumber}
\newcommand{\ph}{\phantom}
\newcommand{\eps}{\epsilon}
\newcommand{\be}{\begin{equation}}
\newcommand{\ee}{\end{equation}}
\newcommand{\bea}{\begin{eqnarray}}
\newcommand{\eea}{\end{eqnarray}}
\newtheorem{conj}{Conjecture}

\newcommand{\plk}{\mathfrak{h}}


\title{How to make a Universe}

\date{}
\author{Paolo M Bassani}
\email{paolo.bassani22@imperial.ac.uk}
\author{Jo\~{a}o Magueijo}
\email{magueijo@ic.ac.uk}
\affiliation{Theoretical Physics Group, The Blackett Laboratory, Imperial College, Prince Consort Rd., London, SW7 2BZ, United Kingdom}

\begin{abstract}
We establish the general conditions under which evolution in the laws of physics and matter creation or destruction are closely intertwined. They make use of global time variables canonically dual to the constants of Nature. Such times flow at a rate determined by what can be interpret as the chemical potential of the fundamental constants (in analogy with phenomenological clocks based on isentropic fluids). The general condition for violations of energy conservation is then that a matter parameter evolves as a function of a gravity clock or vice-versa. This framework can be envisaged as the environment within which 
a natural selection scenario operates, powered by random mutations in the values of the constants of nature (or indeed any other variability in the laws in terms of the times defined above). The prize function is the creation of matter, followed by its preservation. This can be accomplished in an environment where diffeomorphism invariance is among the possible theories, with mutations modelled, for example, on the absorbing Markov chain. In such a set-up the diffeormorphism invariant state with fixed constants (or any nearby state) should be the absorbing state. John Wheeler's ``higgledy-piggledy'' chaotic cosmic start therefore finds a realization in this model, where its own demise and the establishment of order and seemingly immutable laws is also a predection of the model. 
\end{abstract}

\maketitle

\section{Philosophy, natural selection and physics}
One can spend a lifetime discussing the philosophy of the origin of the fundamental laws of Nature, of time, and of the matter in our Universe. Indeed such a discussion is pleasant and should not be forgone. For example, one could contend that matter, laws and time must have emerged together as a single package. For laws to ``appear'' they must have {\it evolved} out of lawlessness, which could itself be defined as extreme {\it variability} in these laws. Thus, time (as a condition for defining variability) comes into the picture, indeed it could be operationally defined via the variability in the laws. In addition,
matter creation (or annihilation) is associated with variability in the laws of physics, as loosely implied by Noether's theorem. It is thus tempting to rehash the ancient Greek cosmology wherein a primordial chaos containing neither rule nor matter gave way to the cosmos of order and fixed, immutable laws, with matter produced as part of the bargain. 
From Hesiod's \textgreek{q'aos}~\cite{Hesiod} to John Wheeler's higgledy-piggledy~\cite{Wheeler,PaulDavis}, there is no shortage of precedents to the idea of order being the daughter of chaos, with concomitant self-generation or even self-reproduction\footnote{The sexed/sexist Jungian view of masculine order and feminine chaos is not in keeping with the ``prefer not to say'' gender option that would probably be selected by the corresponding self-reproducing Gods.}.

As scientists, however, the first question is: how does one mathematically formalize such an idea?  In this paper we will go a fair way towards answering this question. We set up a formalism allowing for a deterministic connection between laws' variability and matter production (or destruction), thereby rigorously implementing the expectation from a loose application of Noether's theorem. Unlike previous work~\cite{evol}, we do not rely on the assumption of homogeneity and anisotropy (of dubious applicability in the higgledy-piggledy phase). A differentiation between matter and gravity, and between what we call local (or rather non-global)  and global variables will suffice.  We then overlay on this framework a statistical model, based on the absorbing Markov chain. This will provide the ``higgledy-piggledy'' aspect of variability. Random evolution will therefore create matter, but as we will show it will also have the ability to take it away. It is therefore important for the successful creation of a Universe that the absorbing state be a state where evolution has ceased and diffeomorphism invariance established,
so that the source, but also the potential sink of matter creation, is turned off. This allows the lucky Universes to keep their gains. 

A close analogy with natural selection therefore arises. The deterministic part of the model  may be seen as the established environment (the objective environmental conditions in a given planet, say). The Markov chain, instead, can be seen as the random  mutation processes providing the engine for natural selection. As in biological natural selection, some random mutations produce Universes with matter, others do not, or worse, produce negative energy/matter. One therefore needs the mutation game to be turned off and stability to establish itself to make sure any possible gains are preserved. Thus, the apparent (near-)constancy of constants and laws we observed, as well as diffeomorphism invariance, are explained in this paper as an evolutionary advantage.

Before jumping into the core of this paper, we sketch a table of contents. In Sections~\ref{environ}--\ref{Section:non-cons} 
we present the ``deterministic environment'' for our evolution model. The basic scaffolding is in Section~\ref{environ} and 
draws heavily on unimodular gravity~\cite{unimod1,UnimodLee1,unimod,alan,daughton,sorkin1,sorkin2,Bombelli,UnimodLee2,JoaoLetter,JoaoPaper} as phrased in~\cite{unimod} and its generalizations~\cite{pad,pad1,lomb,vikman,vikman1,vikaxion,JoaoLetter,JoaoPaper,evol}, and allows for a clear definition of evolution in the laws of physics (inspired by, but going beyond the symmetry reduction explored in~\cite{evol}). 
Concepts we have advertised in this Introduction, such as local and global variables (Sec.~\ref{locglob}), and constants and time as canonical duals (Sec.~\ref{timeasmu}) will play a prominent role and are given proper definitions. 
Time will be defined as a ``chemical potential'' of the constants of Nature in an analogy with phenomenological fluid clocks, where evolution can be seen as a chemical reaction or a non-adiabatic process. 

We then describe two possible features in this environment that will be crucial in the selection process: separation of matter and gravity (Section~\ref{mat-grav}) and diffeomorphism invariance in the local part of the theory (Section~\ref{diffs-cons}). These will be essential in the formula predicting the mass outcome of evolution derived in Section~\ref{Section:non-cons}. 
The rest of the paper is devoted to the ``random mutation'' element of the model and is in Sections~\ref{higgpig1}-\ref{MassFunction}. As explained above this is based on the absorbing Markov chain model.

\section{The basic ``environment'' of evolution}\label{environ}
We start by laying down the formalism needed for supplying the deterministic background to our model of natural selection in Sections \ref{higgpig1} and \ref{higgpig2}. We defer to those Sections the details of where and when randomness appears (dependent on the details of these initial Sections). 
What follows draws on, but significantly expands~\cite{evol,geoCDM,nongeoCDM} (themselves associated with the unimodular work referred above and~\cite{BHevol,MachianCDM}).

\subsection{Local and global variables}\label{locglob}
We assume a preferred foliation, $\Sigma_t$, coordinatized by $(t,x)$. This provides the ``matrix'' for evolution and possibly switching off of the same, with the matrix then being erased from the laws of physics.
Given $\Sigma_t$ we distinguish local and global variables (collectively represented by $\{q(x),p(x)\}$ and $\{\alpha,T_\alpha\}$, respectively) from the structure of the action:
\begin{eqnarray}
     S&=&\int dt\,  V_\infty \dot\alpha T_\alpha +S_L
     \nn\\
     &=&\int dt\,  \left[ V_\infty \dot\alpha T_\alpha
     + \int_{\Sigma_t} d^3x\, (\dot q(x) p(x) -{\cal H}_E[)  \right]
\label{globdef}
\end{eqnarray}
where $V_\infty=\int_{\Sigma_t}  d^3 x$ is the volume of $\Sigma_t$, either finite or with a limiting procedure implied. The global variables are independent of $x$ on $\Sigma_t$, and may be considered a property of the whole leaf $\Sigma_t$. In contrast, each point on the leaf $x$ is associated with canonical pairs of local variables, $q(x)$ and $p(x)$.
The global variables, just as the local ones, are assumed to be {\it intensive} variables, which justifies the factor of $V_\infty$ in the first term of (\ref{globdef}). The extended Hamiltonian ${\cal H}_E$ depends parametrically on the variables $\alpha$ (which will become our constants if there is time independence) and on $T_\alpha$ if there is time dependence or evolution.

The first term in~\eqref{globdef} contains the global variables already in their canonical Hamiltonian form. The global variables are assumed not to have a Hamiltonian of their own. The remaining terms, collected in $S_L$, make up the ``local'' action\footnote{In the sense of non-global; they could included what is usually called non-local in a perturbative sense, i.e. higher order derivative theories and the like. That will not be considered here.}. 
Upon a Legendre transformation of $S_L$, it splits into the local variables' canonical terms and the extended Hamiltonian density, ${\cal H}_E$.  The non-vanishing Poisson brackets therefore are:
\begin{eqnarray}
    \{q(x),p(y)\}&=&\delta(x,y)\\
    \{\alpha,T_\alpha\}&=&\frac{1}{V_\infty}.
\end{eqnarray}
To isolate effects due to global interactions it is useful to define the ``non-local'' or ``global'' Poisson bracket:
\begin{equation}
    \{f,g\}_{NL}=\frac{1}{V_\infty}\sum_{(\alpha ,T_\alpha)}
     \frac{\partial f}{\partial \alpha}\frac{\partial g}{\partial T_\alpha}- \frac{\partial f}{\partial T_\alpha}\frac{\partial g}{\partial \alpha}
\end{equation}
where $f$ and $g$ are general functions of phase space and the sum is over the global variables only.
The full Poisson bracket then splits as:
\begin{equation}
     \{f,g\}= \{f,g\}_{L}+\{f,g\}_{NL}.
\end{equation}

The equations of motion of the theory in Hamiltonian form are:
\begin{eqnarray}
    \dot q(x)&=&\{q(x),{\bf H}\}=\frac{\delta  {\cal H}_E(x)}{\delta p(x)}
    \label{ham1loc}\\
     \dot p(x)&=&\{p(x),{\bf H}\}=-\frac{\delta {\cal H}_E(x)}{\delta q(x)}
\label{ham2loc}
\end{eqnarray}
for the local variables. These are to be 
computed at fixed $\alpha$ and $T_\alpha$, just as one fixes all variables other than the pair involved\footnote{This justifies one of the most ad hoc assumptions in the apparently crude model presented in~\cite{VSL}. Indeed early VSL models are a subclass of the models presented here.}. Hence, the evolution of local variables at a given point only depends on the functional derivatives of the Hamiltonian density ${\cal H}$ at that point.
This is to be contrasted with the equations of motion for the global variables:
\begin{eqnarray}
    \dot\alpha&=&\{\alpha,{\bf H}\}=
    \int_{\Sigma_t} \frac{d^3 x}{V_\infty} \frac{\partial {\cal H}_E}{\partial T_\alpha}
    \label{ham1totb}\\
    \dot T_\alpha&=&\{T_\alpha,{\bf H}\}=-
    \int_{\Sigma_t} \frac{d^3 x}{V_\infty} \frac{\partial {\cal H}_E}{\partial \alpha},
    \label{ham2totb}
\end{eqnarray}
dependent on the Hamiltonian density over the whole leaf, weighted by their fractional volume (leading to the concept of Machian reservoir introduced in~\cite{MachianCDM,geoCDM}). 

\subsection{Time as the chemical potential of the fundamental constants}\label{timeasmu}

One may recognize in the global variables defined above the cosmological constant $\Lambda$ and 4-volume time $T_\Lambda$ as they appear in unimodular gravity~\cite{unimod1,UnimodLee1,unimod,alan,daughton,sorkin1,sorkin2,Bombelli,UnimodLee2,JoaoLetter,JoaoPaper}, specifically as formulated in~\cite{unimod} once the gauge degrees of freedom are eliminated~\cite{geoCDM}. Indeed~\cite{unimod} may be taken as the prototype (with $\alpha=\Lambda$ and $T_\alpha=T_\Lambda$) of a general procedure that can target any fundamental constant or vector of constants, $\alpha$. Examples studied in~\cite{evol} include:
\begin{eqnarray}
    {\bm \alpha}&=&\left(\rho_\Lambda, M_{PL}^2,\frac{G_M}{G_P},c_g^2,c_m^2,... \right),\label{alpha}
\end{eqnarray}
resulting in the well known 4-volume time, Ricci time, Newton time, and others~\cite{pad,pad1,lomb,vikman,vikman1,vikaxion,JoaoLetter,JoaoPaper,evol}.


It is possible to link this procedure (which could at least aspire to being ``fundamental") to more phenomenological but also more tangible procedures for defining time. This comparison provides an interesting physical interpretation: the times $T_\alpha$ flow with a rate that can be seen as {\it the chemical potential of the fundamental constants}. Consider the phenomenological clocks (classical and quantum~\cite{Isham,viqar,gielen,gielen1}) based on an isentropic fluid~\cite{brown,brownkuchar} reduced to minisuperpsace (MSS)\footnote{Such models are ``phenomenological'' because they collapse beyond MSS, whereas our $\alpha$, $T_\alpha$ do not. Perfect fluids of course do not rely on a preferred foliation, but they do not generally provide clocks away from that context.}. 
Recall the equilibrium thermodynamic relation
\begin{equation}
    dU=TdS-pdV+\sum_i\mu_i d\Pi_i
\end{equation}
where $\Pi_i$ are conserved  particle numbers\footnote{Denoted here by $\Pi_i$ instead of the more commonly used in the literature $N_i$, to avoid confusion with the ADM shift function later. As usual $n_i=\Pi_i/a^3$ is the associated MSS density.}. Just as in~\cite{brown} this can be implemented from an action principle, which, reduced to MSS, becomes~\cite{gielen,gielen1}:
\begin{equation}\label{Sfluid}
    S_i=V_c \int dt\left( \dot \Pi_i\Theta_i-Na^3\rho_i\left(n_i\right)\right)
\end{equation}
with $N$ the lapse function and:
\begin{eqnarray}
n_i&=&\frac{\Pi_i}{a^3},\\
    \mu_i&=&\frac{p_i+\rho_i}{n_i}=
    \frac{\partial \rho_i}{\partial n_i},
    \label{mudef}\\
    p_i&=&n_i\frac{\partial\rho_i}{\partial n_i}-\rho_i\label{pdef}
\end{eqnarray}
where $n_i$ is the density associated with the conserved number and $\mu_i$ and $p_i$ are the chemical potential and pressure. The Hamilton equatio for $\Theta_i$ implies that it can be used as a clock ticking at a rate defined by the chemical potential of the fluid:
\begin{align}
    \dot \Theta_i&=-N\mu_i\label{dotTheta},
\end{align}
with the ``time independence'' of the Hamiltonian enforcing the conservation law  for its dual number density, $ \dot \Pi_i=0$. For an equation of state $p_i=w_i\rho_i$ with $w_i\neq -1$, we have 
\begin{equation}
\rho_i\propto n_i^{1+w_i}.     
\end{equation}
For dust, $\rho=mn$, with $m$ the rest mass of its grains, so the chemical potential is a constant $\mu=m$, resulting in its phenomenological time being proportional to the proper time felt by the dust grains. For radiation the number of photons in a comoving volume is conserved, but due to redshift $\rho\propto n^{4/3}$.  The chemical potential is proportional to the temperature, so that the resulting time is proportional to conformal time.

The MSS fluid reduction \eqref{Sfluid} mimics the expression \eqref{globdef} (which is valid beyond MSS) with the dictionary:
\begin{align}
    \rho_i&\rightarrow \frac{\mathbf{H}}{V_c}=\frac{1}{V_c}\int_{\Sigma_t}d^3x\, {\cal H}_E \label{anal1}\\
    \Pi_i&\rightarrow \alpha\label{anal2}\\
    \Theta_i&\rightarrow T_\alpha.\label{anal3}
\end{align}
Thus, in the fluid analogy, the constants of nature parallel the conserved particle numbers of isentropic fluids, and the time dual to the constants mimics the Lagrange multiplier $\Theta_i$ enforcing their constancy via absence of time dependence. 
By comparing  \eqref{ham1totb} with \eqref{dotTheta}) 
we see that the rate of flow of time $T_\alpha$ can then be interpreted chemical potential of the constant, defined as:
\begin{equation}
   \mu_\alpha= \frac{1}{V_c}\frac{\partial \mathbf{H }}{\partial \alpha}
\end{equation}
replacing $\mu_i$ and completing our analogy.

Evolution in this analogy is therefore akin to a chemical reaction, or any other situation in which the usually conserved particle number is made to change, such as non-adiabatic evolution. It results in a change of the fundamental constants dual to the times ruling the change.

\section{Separation of matter and gravity}\label{mat-grav}

Within the ``chaos" there is a priori no reason for why matter and gravity should be separated\footnote{As a footnote one can't help wondering what the implications for a theory of quantum ``gravity'' might be, given that ``quantum'' gravity might only have any relevance in a phase where gravity and matter would not be separated.}. In what follows we show that, even with the most modest prejudices, this separation leads to an evolutionary advantage. Towards the end of this paper, we will suggest a quantitative measure of such separation. We start by providing a general definition. 

The basic requirement is that there be separate gravity and matter local and global variables, as well as separate terms in the local action 
\begin{equation}
    S_L=S_M+S_G.
\end{equation}
Gravity should be associated with geometry, for example by containing the metric among its variables. Two components of the metric, the lapse $N$ and the shift $N^i$, distinguish themselves by requiring that the
extended Hamiltonian density contain a Hamiltonian and momentum constraints:
\begin{equation}
    {\cal H}_E=N{\cal H}+N^i{\cal H}_i+...
\end{equation}
with a similar split:
\begin{equation}
    \mathbf{H}_E=\mathbf{H}+\mathbf{P}+...
\end{equation}
for their integrated {\it global} equivalents: 
\begin{align}
{\bf H}_E&=\int_{\Sigma_t} d^3 x\, N {\cal H}_E(x)\\
  {\bf H}\equiv H(N)&=\int_{\Sigma_t} d^3 x\, N {\cal H}(x)\\
  \mathbf{P}\equiv H_i(N^i)&=\int_{\Sigma_t} d^3 x\, N^i {\cal H}_i(x). 
\end{align}
Given the split in $S_L$ the constraints split as:
\begin{eqnarray}
     {\cal H}&=&{\cal H}_M+{\cal H}_G\\
      {\cal H}_i&=&{\cal H}_{i}^M+{\cal H}_{i}^G§,
\end{eqnarray}
with a similar split of the global terms. 

We assume that matter couples to gravity, and so it must contain the metric degrees of freedom. The dependence is encapsulated in the energy-momentum tensor:
\begin{equation}
     T^M_{\mu\nu} =
    \frac{-2}{N\sqrt{h}}\frac{\delta S_M}{\delta g^{\mu\nu}}.
\end{equation}
where $h$ is the determinant of the spatial metric $h_{ij}$ on $\Sigma_t$. 
In order to isolate the issues of matter creation by time evolution we will 
assume that the local theory (at fixed global variables, or taking only the local Poisson brackets) satisfies:
\begin{align}\label{NLWEP}
     \left.\nabla_\nu T^{\mu\nu}_M\right|_{L}&=0. 
\end{align}
This may be seen as a general definition of the equivalence principle in the presence of evolution/global variables. Any violations will therefore arise from the non-local terms in the Poisson brackets, focusing our discussion on the effects of evolution\footnote{In practice we are assuming that $S_G$ does not contain the matter degrees of freedom, avoiding confusing matters with local fifth forces.}.  

\section{Implicatins of diffeomorphism invariance of the local theory}\label{diffs-cons}

Another crucial evolutionary feature is
the diffeomorphism invariance of the theory in the absence of global interactions (and so, of evolution). This helps satisfy the condition~\eqref{NLWEP} (which is stronger). It amounts to requiring that the local Poisson bracket terms of the algebra of constraints satisfy the Dirac hypersurface deformation algebra (HDA):
\begin{align}
    \{H_i(N^i), H_j(M^j)\}_L&= H_i([N,M]^i)\label{smearhihi}\\
    \{H_i(N^i), H(N)\}_L&= H(N^i\partial _iN)\label{smearhih0}\\
    \{ H(N), H(M)\}_L&= H_i(h^{ij}(N\partial_j M- M\partial_j N)).\label{smearh0h0}
\end{align}
These imply the generalized equivalence principle \eqref{NLWEP} with extra assumptions on the coupling of matter to gravity (see~\cite{nongeoCDM}). They will also lead to a nice formula for the violations of \eqref{NLWEP} due to evolution. 

The starting point is the  time evolution of the Hamiltonian and momentum (or the off-shell evolution in standard theories). It was found in~\cite{geoCDM,nongeoCDM} that their free evolution (i.e. in the absence of global terms) under the HDA is given by: 
\begin{align}
   \dot {\cal H}|_L=\{{\cal H},\mathbf{H}\}_L&=\partial_i(N^i {\cal H})+\partial_i ({\cal H}^i N)+ {\cal H}^i \partial_i N \label{dotHH}\\
        \dot {\cal H}_i|_L=\{{\cal H}_i,\mathbf{H}\}_L
        &= {\cal H}\, \partial_i N +\partial_j(N^j {\cal H}_i)  + {\cal H}_j\partial _i N^j.
        \label{dotHHi}
\end{align}
The evolution of the {\it matter} Hamiltonian and momentum is more complicated and relies on extra assumptions. It is known (e.g., \cite{brownkuchar} Sec.~IVA, or~\cite{nongeoCDM}) that diffeomorphism invariance does {\it not} imply that the matter Hamiltonian and momentum satisfy separately the HDA, i.e. s version of \eqref{smearhihi}-\eqref{smearh0h0}. Rather, this algebra is only valid for the {\it total}  ${\cal H}$ and ${\cal H}_i$ as well as for their gravity contributions. Non-vanishing cross terms between gravity and matter therefore spoil the algebra for matter. This happens fundamentally because gravity is always present and is non-negotiable, whereas matter is optional. As a result, if there is diffeomorphism invariance, gravity satisfies HDA on its own.  

A detailed discussion 
is not needed in this paper. 
Here we will just write the evolution as:
\begin{align}
       \dot {\cal H}_M&= \{{\cal H}_M, {\mathbf{H}}\}=
      \{{\cal H}_M, {\mathbf{H}}\}_{Lwep}+ \{{\cal H}_M, {\mathbf{H}}\}_{NL},\label{dotHHmtot}\\
       \dot {\cal H}^i_M&= \{{\cal H}_M^i, {\mathbf{H}}\}=
      \{{\cal H}^i_M, {\mathbf{H}}\}_{Lwep}+ \{{\cal H}^i_M, {\mathbf{H}}\}_{NL}.\label{dotHHimtot}
\end{align}
the $wep$ subscript indicating that the extra terms with respect to \eqref{dotHH} and \eqref{dotHHi}  are subject to assumptions on the coupling to gravity leading to \eqref{NLWEP}. We then decompose:
\begin{align}
      T^M_{\mu\nu}&=\rho n_\mu n_\nu-( n_\mu\Pi_\nu +   n_\nu\Pi_\mu)+\Pi_{\mu\nu} 
\end{align}
with:
\begin{align}
    \rho&=n^\mu n^\nu T^M_{\mu\nu}=\frac{{\cal H}_M }{\sqrt{h}}\\
      P_\mu&=T^M_{\mu\nu}n^\nu=\frac{1}{\sqrt{h}}(N{\cal H}_M +N^k{\cal H}^M_k,{\cal H}^M_i)\\
    \Pi^\mu&= h^{\mu\alpha}n^\nu T^M_{\alpha\nu}=h^{\mu\alpha}P_\alpha={\left(0,\frac{{\cal H}^i_M}{\sqrt{h}}\right)}\\
    \Pi_{\mu\nu}&=h_\mu^{\; \alpha} h_\nu^{\; \beta} T^M_{\alpha\beta}
\end{align}
as well as the associated:
\begin{align}
      P^\mu&=\frac{1}{{\sqrt{-g}}}(-{\cal H}_M, N^i{\cal H}_M+N{\cal H}_M^i)\\
         \Pi_\mu &=\frac{1}{\sqrt{h}}(N^k{\cal H}^M_k,{\cal H}^m_i)
\end{align}
where $n_\mu =(-N,0)$ is the normal 
and $h_{\mu\nu}=g_{\mu\nu}+n_\mu n_\nu$ the projector associated with $\Sigma_t$, defined as in the standard ADM formalism. 
From assumption \eqref{NLWEP} the result will necessarily be:
\begin{align}
     n_\mu\nabla_\nu T^{\mu\nu}_M&=\frac{1}{\sqrt{-g}}\left(-\dot {\cal H}_M +\{{\cal H}_M,\mathbf{H}_E\}_{Lwep}\right)\nn\\
       h_{i\mu}\nabla_\nu T^{\mu\nu}_M&=\frac{1}{\sqrt{-g}}\left(-\dot {\cal H}^i_M +\{{\cal H}^i_M,\mathbf{H}_E\}_{Lwep}\right),\nn
\end{align}
so that inserting \eqref{dotHHmtot} and \eqref{dotHHimtot} finally leads to:
\begin{align}
        n_\mu\nabla_\nu T^{\mu\nu}_M&=-\frac{1}{\sqrt{-g}} \{{\cal H}_M,\mathbf{H}_E\}_{NL}\label{nDiv}\\
          h_{i\mu}\nabla_\nu T^{\mu\nu}_M&=-\frac{1}{\sqrt{-g}} \{{\cal H}^M_i,\mathbf{H}_E\}_{NL}.\label{hDiv}
\end{align}
We will {\it assume} that the RHS of \eqref{hDiv}, but not that of \eqref{nDiv}, is zero (this could be generalized). Hence, non-local interactions produce energy, but not momentum, on the preferred foliation $\Sigma_t$. Note that $N=N(t)$ while global interactions are taking place, since the full diffeomorphisms are broken down to spatial ones. Hence, there is a global Hamiltonian constraint but not a local one. In contrast, $N^i=N^i(x,t)$, so there is always a local (and so also a global) momentum constraint. That is:
\begin{align}
    {\mathbf{H}}&=0\\
    {\cal H}_i&=0.
\end{align}
None of these constraints applies separately for matter and gravity.

\section{Matter production and evolution}\label{Section:non-cons}
We thus arrive at a generalization of the central formula in~\cite{evol} relating matter production and evolution. Notice that it is possible for $ \{{\cal H}_M,\mathbf{H}_M\}_{NL}\neq 0$. However, upon integration over $\Sigma_t$ this term disappears and we are left with:
\begin{align}
      -\int_{\Sigma_t} d^3 x\, \sqrt{h }\, n_\mu\nabla_\nu T_M^{\mu\nu}= \frac{1}{N(t)}\{\mathbf{H}_M,\mathbf{H}_G\}_{NL}\nn\\
      =\frac{1}{N(t)} \sum_{I }
     \frac{\partial {\mathbf H}_M}{\partial \alpha_I}\frac{\partial {\mathbf H}_G}{\partial T_{\alpha I}}-
\frac{\partial {\mathbf H}_M}{\partial T_{\alpha I}}\frac{\partial {\mathbf H}_G}{\partial \alpha_I}. 
\label{totnoncons1}
\end{align}
This is generic but we can make it more concrete by assuming that the time dependence happens via a set of parameters that become fundamental constants when the variability stops. To avoid second class constraints, there should not be any overlap between such ``constants'' and those used to define time. Specifically, we should select a number of target constants split as:
\begin{equation}
    \bm\gamma=({\bm{\alpha}},\bm{\beta}).
\end{equation}
Their fate is to {\it either} become an $\bm \alpha$  (generator of the dual clocks $T_{\alpha I}$) {\it or} become a function:
\begin{equation}
     \bm\beta= \bm\beta ({\bm T}_{\bm\alpha})
\end{equation}
which can be seen as potentials encoding the evolution of the laws of physics. 
Then, \eqref{totnoncons1} becomes:
\begin{widetext}
\begin{eqnarray}\label{centralformula}
    -\int_{\Sigma_t} d^3x \sqrt{h}\,  n_\mu \nabla_\nu T^{\mu\nu}
    &=& \frac{1}{N(t)} \sum_{I K}
\frac{\partial \beta_K }{\partial T_{\alpha I}}\left(
     \frac{\partial H_M}{\partial \alpha_I}\frac{\partial H_G}{\partial \beta_K}-
\frac{\partial H_M}{\partial \beta_K}
\frac{\partial H_G}{\partial \alpha_I}\right) \nn 
\end{eqnarray}
\end{widetext}
For FRW this reduces to the central result Eq.82 in~\cite{evol} (once one corrects a minus sign typo, and adjusts the notation regarding $V_c$ factors). But this expression is more general and only relies on the existence of a foliation $\Sigma_t$. 
Given this foliation the generic formula for matter creation takes the form:
\begin{align}\label{dotMgeneral}
    \dot M=\frac{\partial \beta_K }{\partial T_{\alpha I}}M^{IK}
\end{align}
(with Einstein summation convention applied to the matrix indices), where $M$ is the mass on $\Sigma_t$, $M_{IJ}$ is the mass production matrix for a given local theory and given a choice of $ \bm\gamma=({\bm{\alpha}},\bm{\beta})$, and the first factor on the right hand side encodes the varaibiliy predicted by the potentials $ \bm\beta= \bm\beta ({\bm T}_{\bm\alpha})$. 

\section{Levels  of Higgledy-piggleding }\label{higgpig1}
So far we have developed a deterministic framework for evolution in the laws of physics. It contains a 
deterministic mechanism relating evolution and matter production (or destruction). Thus, we have a concept of ``mutation'' and how it can lead to positive or negative outcomes (matter creation or destruction). 
But how do we make such mutations random, as in  more traditional natural selection? And given that we want order to emerge from ``total chaos'', how much is it meaningful to specify about such random process? 

At the most basic level the potentials ${\bm\beta}({\bm T}_{\bm\alpha})$ should be random functions. For example they could be a series of step functions with random steps.  ``Random'' should be so to the point of the distribution being irrelevant or itself random (for example in a form of recursive, iterated Dirichlet process~\cite{dirichlet}). As it happens, these details do not matter, because whatever ``randomness'' we assign to such 
random steps,  \eqref{dotMgeneral} transmutes into the generic structure:
\begin{equation}\label{mass1}
    \delta M=\sum_k\Delta^k_{IJ}J(k)M^{JI} 
\end{equation}
where $k$ indices the step, $\Delta^k_{IJ}$ is the value of the step in $\beta_I(T_{\alpha J})$, and $J(k)$ encodes the time dependent Jacobian needed to convert a jump in 
$T_{\alpha J}$ into a jump in $t$ as well as other time dependence: it may be an increasing or a decreasing function, depending on the specific model. This is enough for predicting positive or negative outcomes (mass production or destruction). 

At a deeper level we could consider the Hamiltonian itself to be a random function of both the local and global variables, resulting in random matrices $M^{IK}$. Then, some of these theories have {\it local} diffeomorphism invariance (only broken by the global variables, as we have considered here) others do not (see the implications this has on this discussion in~\cite{HLPaolo}). Taking this into account, we can describe the general process as a marginalization of conditional processes satisfying \eqref{mass1}. 
But the point is that this added layer of randomness is not part of natural selection, but, rather, is similar to the randomness that renders some planets fit for life and others not, obviously through a process which contains a lot of chance, but which is a different random process with regards to the natural selection process leading to life.

Therefore let a  ``planet'' be a system with a local Hamiltonian and global variables $\{{\bm \alpha}, {\bm T}_{\bm \alpha}\}$, as well as a set of target ${\bm \beta}$ which can depend on ${\bm T}_{\bm \alpha}$.  Let us take a ``planet'' with $M_{IJ}\neq 0$ (which defines and quantifies separation of matter and gravity).  If we know nothing about the random process ruling $\Delta$, or even if we deny a meaningful sense to such knowledge, there should not be any bias in its sign. Hence matter can be equally created or destroyed.  Therefore, in order to have the chance of gains and of keeping them we need three things:
\begin{itemize}
\item That the random Hamiltonians contain Hamiltonians with a clear separation of matter and gravity, in the sense of $M_{IJ}\neq 0$ so that the conditionalized random steps can generate matter. 
    \item That the random Hamiltonians contain Hamiltonians with local diffeomorphism invariance, so that we can conditionalize our random processes to those. 
    \item  That we stop the random mutations within that subset (this is not a requirement in some cases, as we shall see later, but it certainly ensures success). 
\end{itemize}
In this sense, near-diffeomorphism invariance is a low mutation regime (like the low radiation environment we live in, as opposed to the primordial Earth), needed to keep the gains of chaos. We are the lucky ones who stopped gambling after a good run. The outcome is a world with a clear separation of matter and gravity, and with near diffeomorphis invariance; as well as the matter that was produced in the previous phase.

We now provide a concrete model for implementing random natural selection conditional to such a ``planet''.

\section{A concrete model for cosmic natural selection }\label{higgpig2}

In a homogeneous absorbing Markov process~\cite{markov}, all states can reach an absorbing state that, once entered, cannot be left. This can model the shift of a process from a purely stochastic to a deterministic one. 
In such processes, the transition matrix is of the form:
\begin{equation}
\begin{bmatrix}
Q & R \\
\bm{0} & \bm{I} 
\end{bmatrix}, \label{gen_abs}
\end{equation} 
where $Q$ is the $n \times n$ transition matrix between transient states, $R$ is the $n \times m$ transition matrix between transient and absorbing states, $\bm{0}$ is a $m \times n$ zero matrix and $\bm{I}$ is the identity matrix (since once an absorbing state is entered, there is always zero probability of transitioning to any other state). We will now define one such model adapted to our natural selection requirements. 

\subsection{Evolution and Universes}

In our model,  the $\bm{\beta}$ potentials are step functions, such that at each Markov iteration a value for the step function is chosen randomly. In general, these can be both negatively and positively valued steps, such that matter can be equally created and destroyed. To avoid biases towards more positively or negatively valued steps overall, the absorbing state is moved to the center of the matrix, modifying \eqref{gen_abs} to:
\begin{equation}
T = 
\begin{bmatrix}
p_{11}& \dots & \alpha_1 & \dots&p_{1j}  \\ 
\vdots & \ddots & \alpha_2& \ddots&\vdots\\ 
0& 0& 1& 0&0\\  
\vdots & \ddots& \alpha_3& \ddots& \vdots\\ 
p_{i1} & \dots& \alpha_n&\dots& p_{ij} 
\end{bmatrix},
\end{equation}
where $\alpha_n$ is the probability of transitioning from any given transient state to the absorbing one. The transition probabilities are drawn randomly from a uniform distribution, including for the $\alpha_n$.
In this new ``Templar'' representation, $T$ is an $n \times n$ 
matrix, where the Markov states' values are $S = [-\infty, \dots, 0, \dots, +\infty]$. Additionally, $n$ must be odd so that the absorbing state not only is the median value $S_{(n + 1)/2}$, but also coincides with the value $0$, avoiding any biases towards positive or negative steps. Therefore, all the columns before the absorbing state $S_{abs}$ are negative steps, while all the others after are positive ones. 

While $T$ should be infinite to account for all the possible values a given $\bm{\beta}$ visits during evolution, in the following simulation, for simplicity, we use an $11 \times 11$ transition matrix. Furthermore, different $\bm{\beta}$ could, in principle, follow different transition probabilities, but in this example we restrict ourselves to one constant only.  The steps' values are equally spaced integers such that $S = \{-5, \dots, 0, \dots, 5\}$, where the absorbing state is the value $0$, the negative values correspond to $T$'s indices $[0, 4]$, while the positive ones are in the range $[6, 10]$. As an illustrative example, in Figs.~\ref{figg1} and~\ref{fig1} we consider $15$ different realizations of the Markov chain produced using $T$, where the starting state is picked randomly from $S$ with uniform probability. The total matter produced by evolution (hereforth called mass contribution) is quantified by equation \eqref{mass1}, which, in our model, takes the form:
\begin{equation}
    M(n, k) = \sum_k (n_k - n_{k-1})f(k), \label{ploteqn}
\end{equation}
where $n \in S$ and $k$ is the Markov iteration index. The function $f(k)$ represents the matrix $M^{IJ}$ and the Jacobian $J$ and all other time dependent factors defined after~\eqref{mass1}. 
A crucial qualitative difference for the final result is whether $f(k)$ is an increasing or a decreasing function. 
Here, for illustrative purposes we shall investigate the form:
\begin{eqnarray}
    f(k) &=& \frac{k_0}{k_0 + k} \label{one} 
\end{eqnarray}
(where $k_0$ is a modulating parameter) and:
\begin{eqnarray}
    f(k)=k_0 k
\end{eqnarray}
for an increasing function. 


\subsubsection{Decreasing modulating functions $f(k)$}
\begin{figure}
    \centering
    \includegraphics[width=1.05\columnwidth, height=6.7cm]{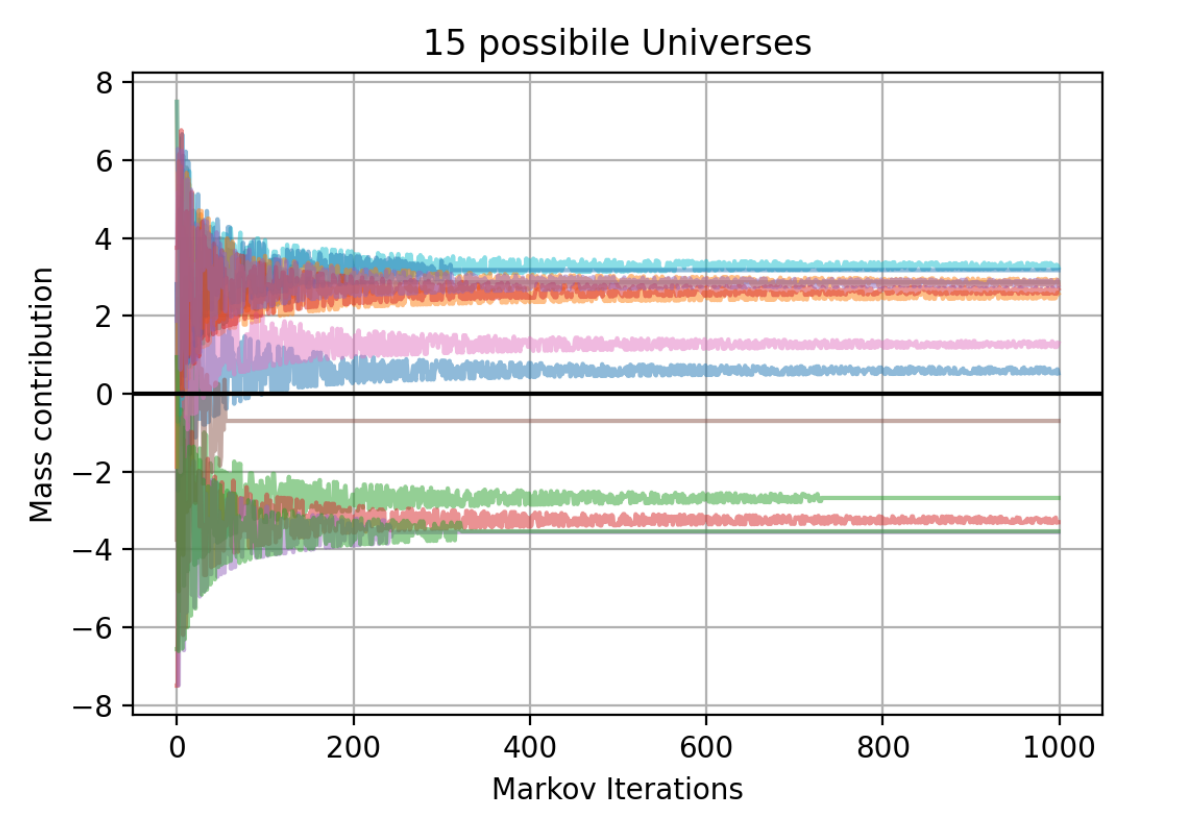}
    \caption{The cosmic mass as it evolves in  $15$ realizations of absorbing Markov chains with 1000 steps, choosing \eqref{one} with $k_0 = 15$. Positive and negative mass Universes are clearly separated.}
    \label{figg1}
\end{figure}
\begin{figure}
    \centering
    \includegraphics[width=1.05\columnwidth, height=6.7cm]{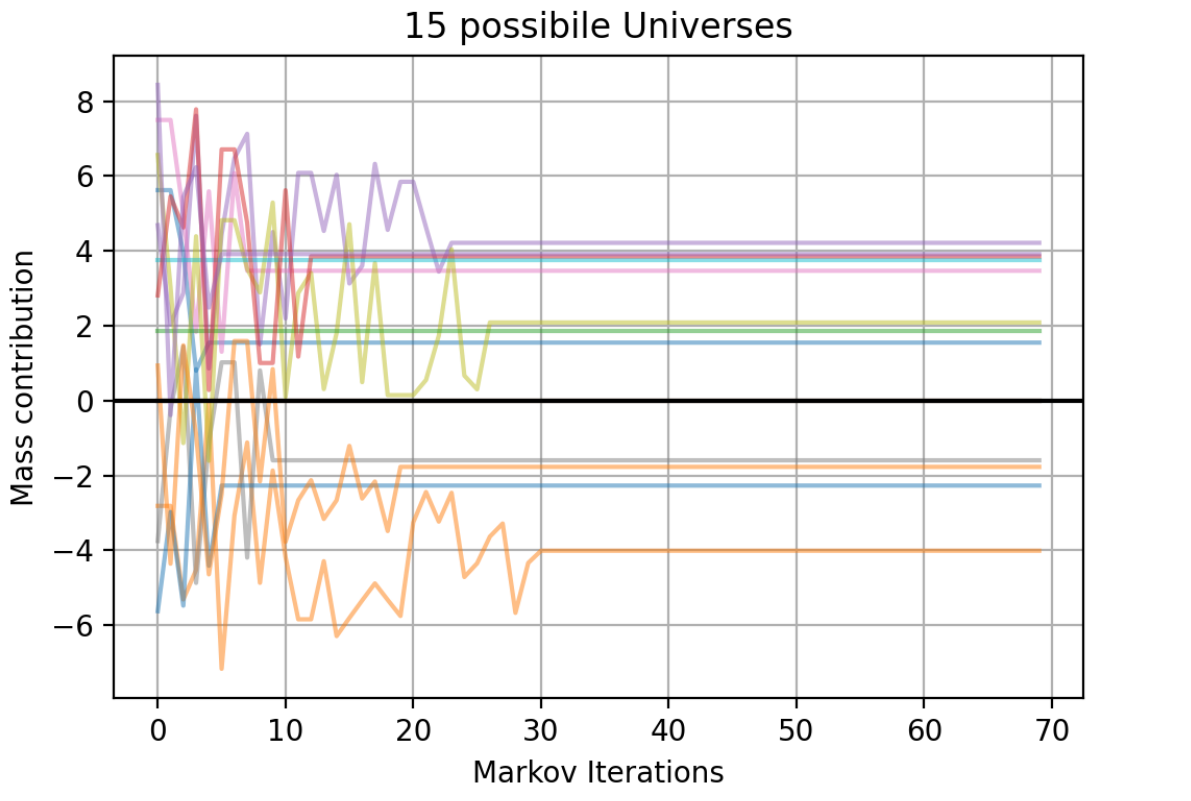}
    \caption{
    The cosmic mass as it evolves in 
    $15$ realizations of absorbing Markov with a matrix $T$ where the transition probabilities into the absorbing state $\alpha_n$ are of the same order as the other $\mathcal{O}(p_{ij})$ (here $k_0=70$).}
    \label{fig1}
\end{figure}

Within this framework, equation \eqref{ploteqn} combined with function \eqref{one} generates typical realizations as depicted in Fig.~\ref{figg1}, should the function $f(k)$ be decreasing. Each line follows the mass of the Universe as it evolves along a possible Markov chain, in a process where the transition probabilities to the absorbing state are two orders of magnitude smaller then the others. 
For a decreasing $f(k)$ the mass gains gathered early on dominate.
This results in the Universes' masses being sharply divided between negative and positive, as determined by the first step.

Natural selection does the rest: as pointed out in ~\cite{evol}, only gravitational parameters depending on matter clocks or vice-versa produce matter, leaving the other combinations sterile. While in this example we have used $f(k)$, implementing the full $M^{IJ}$ from equation \eqref{totnoncons1} would effectively select only some Universes over others, acting as a deterministic mechanism of natural selection between Universes. Universes with negative masses would go extinct and the process will continue until converging on the Universe we live in. In a very Leibnizian spirit, we could say that we actually live in the best of all possible worlds. 


As another example we consider a process where 
the $\alpha_n$ are of the same order of magnitude as the other transition probabilities. This is illustrated in Fig.~\ref{fig1}, with parameter
$k_0 = 70$.
In this example, fewer iterations are required before all realizations reach the absorbing state. 
It can be clearly seen that the Universes' masses undergo random oscillations early on in the chain, eventually settling on a constant mass. 

Trying out other variations we find that  
the lower the $\alpha_n$ with respect to the $p_{ij}$, the higher the number of iterations before the majority of realizations enter the absorbing state. 

\subsubsection{Increasing modulating function $f(k)$}
\begin{figure}
    \centering
    \includegraphics[width=1.05\columnwidth, height=6.7cm]{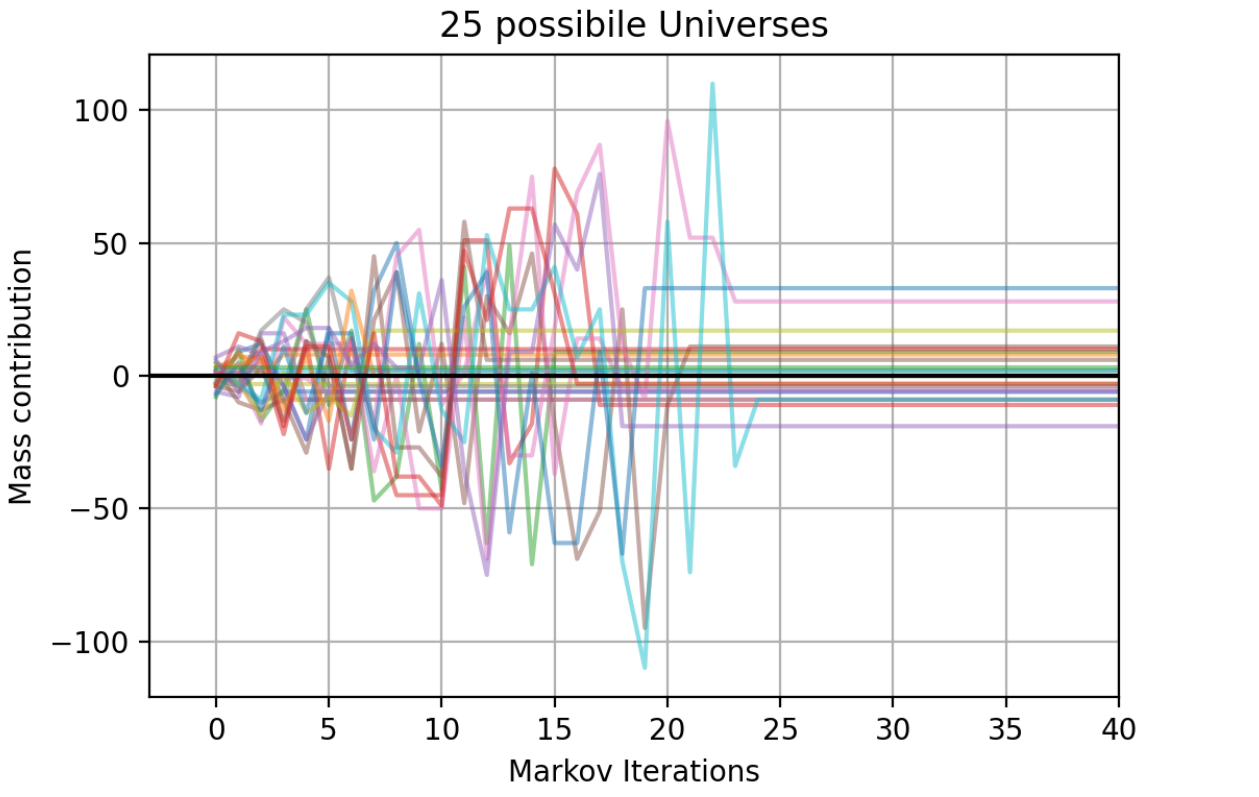}
    \caption{
    The cosmic mass as it evolves in 
    $25$ realizations with an increasing $f(k)=k$. The jumps in mass increase along the chain, with frequent changes of sign. }
    \label{fig2}
\end{figure}

Qualitatively different results are obtained if $f(k)$ increases. Then, in contrast with the last examples, the last steps before hitting the absorbing state dominate and there are frequent swaps in the sign of the mass along the chain. This is illustrated in Fig.~\ref{fig2}. If there were no absorbing state the fluctuations in mass would continue to grow, and any gains would be potentially lost in subsequent iterations.

It is also possible for the function $f(k)$ to be non-monotonic, with in increasing and decreasing pieces, in which case the behaviour of the chains will be a combination of the two listed above, with the outcome depending on the probability of hitting the absorbing state (i.e. how far along one must go to hit it, and so in what regime one is, at that point). 

Note that strictly speaking the Universe can keep its gains even without an absorbing state if $f(k)$ decreases sufficiently fast at late times, even if it was increasing at early times. However, just like we should not assume a bias towards positive mass in our Markov chains, we should not assume that $f(k)$ is forever decreasing until the absorbing state is reached. Hence the need for an absorbing state even if $f(k)$ is piece-wise decreasing. 

\subsection{Possible model refinements}
Two important remarks regarding the absorbing state and the transition matrix are in order. First, this model could be improved by introducing the concept of an absorbing {\it band}. Instead of having a single absorbing state, a limited range (possibly also continuous) of states where variation happens without ever exiting the boundaries can be implemented. Such design would modify the transition matrix $T$ to:
\begin{equation}
T_{band} = 
\begin{bmatrix}
p_{11}& \dots & \alpha_1 & \dots&p_{1j}  \\ 
0& \omega_{11} & \omega_{*}& \omega_{1\nu}& 0\\ 
0& \vdots& \alpha_{*}& \vdots&0\\  
0& \gamma_{\mu1}& \gamma_{*}& \gamma_{\mu \nu}& 0\\ 
p_{i1} & \dots& \alpha_n&\dots& p_{ij} 
\end{bmatrix}, \label{bandt}
\end{equation}
where $\omega$ and $\gamma$ denote the upper and lower boundaries of the band and $\omega_{*}$ and $\gamma_{*}$ are the probabilities of transitioning to the center of the band. Hence, the absorbing band is defined by a nested $m \times m$ transition matrix, whose entries are the $\omega_{\mu \nu}$ and $\gamma_{\mu \nu}$ and where the states between the boundaries could be continuous. This could have important phenomenological implications, since it would allow the constants $\bm{\beta}$ and $\bm{\alpha}$ to jitter even nowadays. The case of the fine structure constant springs to mind~\cite{Murphy:2003hw, Webb,newWebb}. This will be studied further in~\cite{markov2.0}.

Second, in the name of true randomness, there is no reason why we should restrict ourselves to a single transition matrix. There could be multiple transition matrices for the different $\bm{\beta}$, the transition probabilities could be time-dependent and the absorbing band's structure could change from constant to constant. 

\section{The mass distribution function}\label{MassFunction}

\begin{figure}
    \centering
    \includegraphics[width=1.05\columnwidth, height=6.7cm]{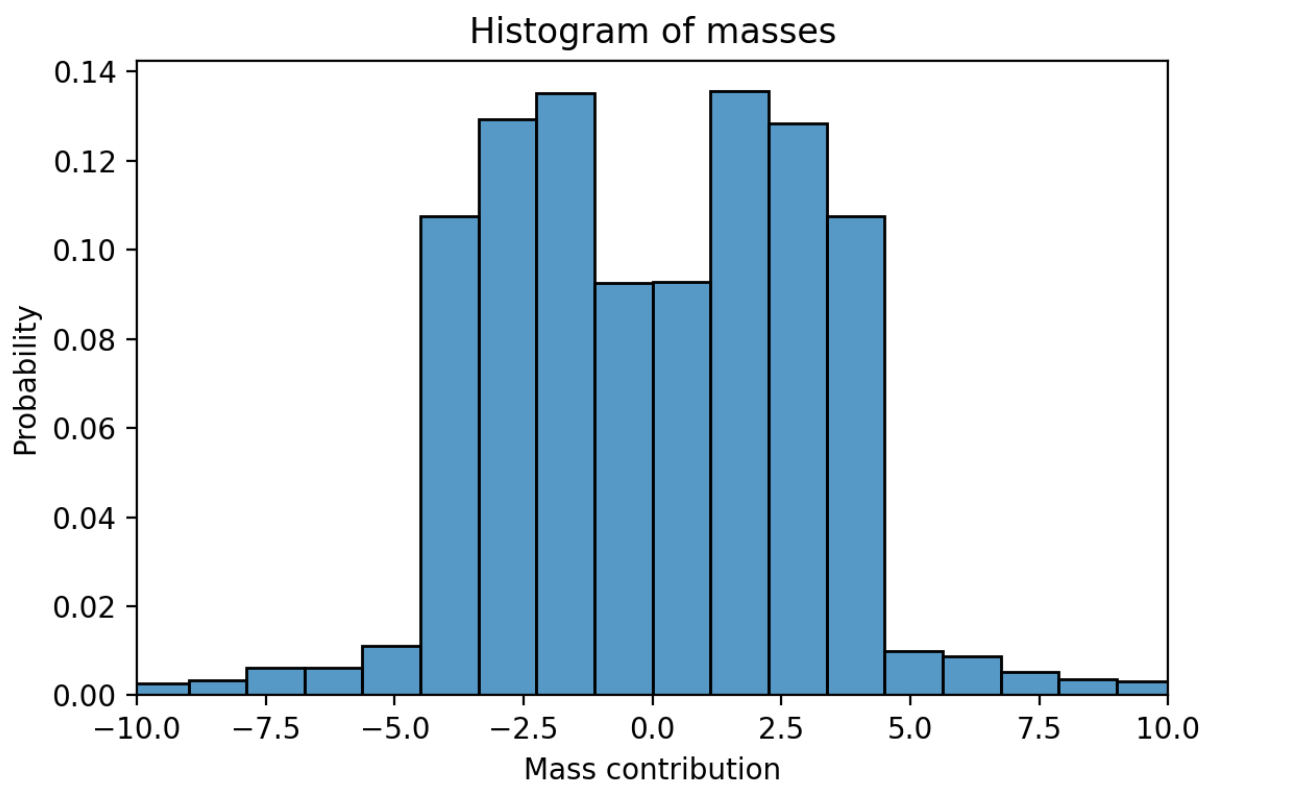}
\includegraphics[width=1.05\columnwidth, height=6.7cm]{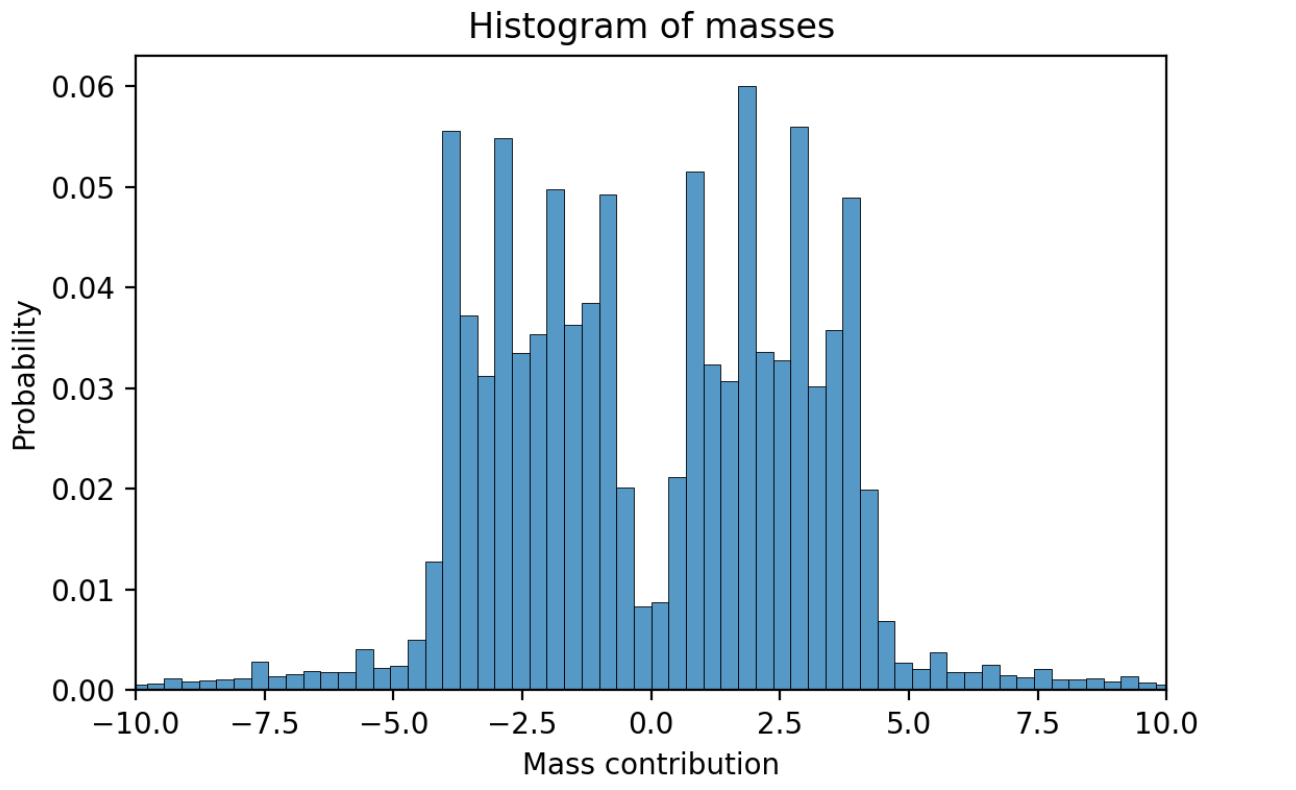}
    \includegraphics[width=1.05\columnwidth, height=6.7cm]{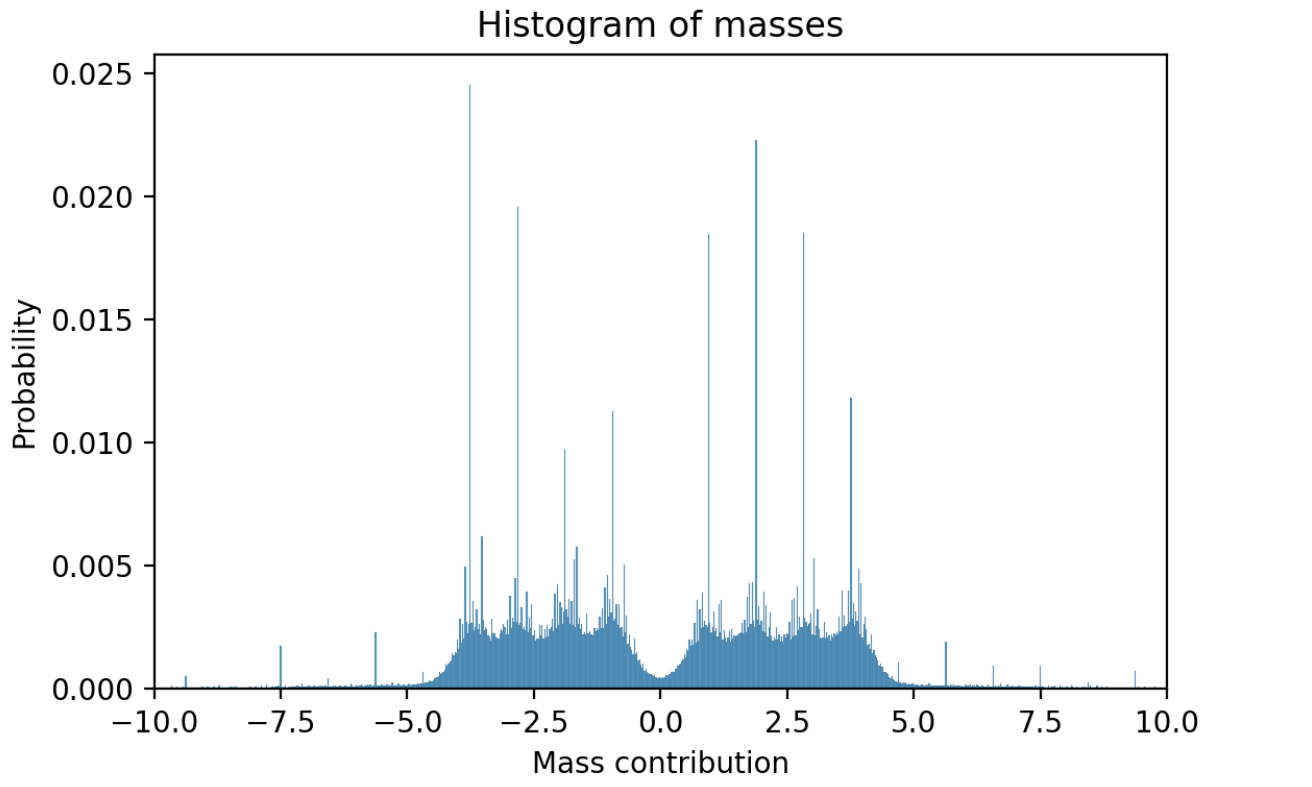}
    \caption{Histograms of the mass accrued by the time the absorbing state is reached, for an 11 state system as described in the text. We used $30$ (top), $100$ (middle) and $621$ (bottom) bins to build these histograms, revealing the fractal structure of peaks explained in the main text.}
    \label{h1}
\end{figure}

We finally estimate the distribution function for the mass outcome of the absorbing Markov chains defined in the previous Section. This has an element of novelty and a mathematical interest in its own right. Most Markov chain studies focus on the distribution of the number of times a given intermediate state has been visited. The  distribution of the path integral over the chain is usually irrelevant. For our ``cosmic selection'' purposes, however, the precise opposite is true. We do not care about how many times a given value of the constants has been visited during the higgeldy-piggledy phase; only whether or not a suitably amount of matter was accrued over the chain by the time the absorbing state is reached. Due to its intrinsic interest we will include below some results that are not necessarily relevant in a cosmological setting, e.g. chain integrals with a small number of possible states (exacerbating the effects of discretization, which may or may not apply to our case).

We estimated the mass probability function by Monte Carlo simulation for various models based on last Section.
A common feature to all histograms is a ``delta''-like peak in the probability of mass centered at zero. This is driven by the Universes whose evolution never got off the ground: the first iteration of their Markov process starts from the absorbing state, making the process, \textit{de facto}, fully deterministic. Because these Universes never departed from the orthodox paradigm of fixed physical laws, we call them good sons, contrasting them with the Universes that did leave constant physical laws behind, just to come back later on, as remorseful prodigal sons. For clarity we remove this delta function peak from most of the histograms we will present (they tend to distort the scale).

Once this is done, we find an interesting feature in models where the number of possible states is small (exaggerating discretization features that may or may not be relevant in a cosmological setting, but which might be interesting in Markov chain theory). We find that the histograms have a ``fractal''-like property: a structure of peaks which survives re-binning revealing thinner/sharper scales. 
This is demonstrated in Fig.~\ref{h1} with histograms of the final mass, derived from 1 million Markov chains evolved long enough so as to reach the absorbing state. We chose the 11 state matrix introduced in the previous Section, and $f(k)$ as in~\eqref{one} with $k_0=15$. We then chose different binnings, the peak structure depending on how coarse or sharp we want the resolution of the histograms to be.

\begin{figure}
    \centering
    \includegraphics[width=1.05\columnwidth, height=6.7cm]{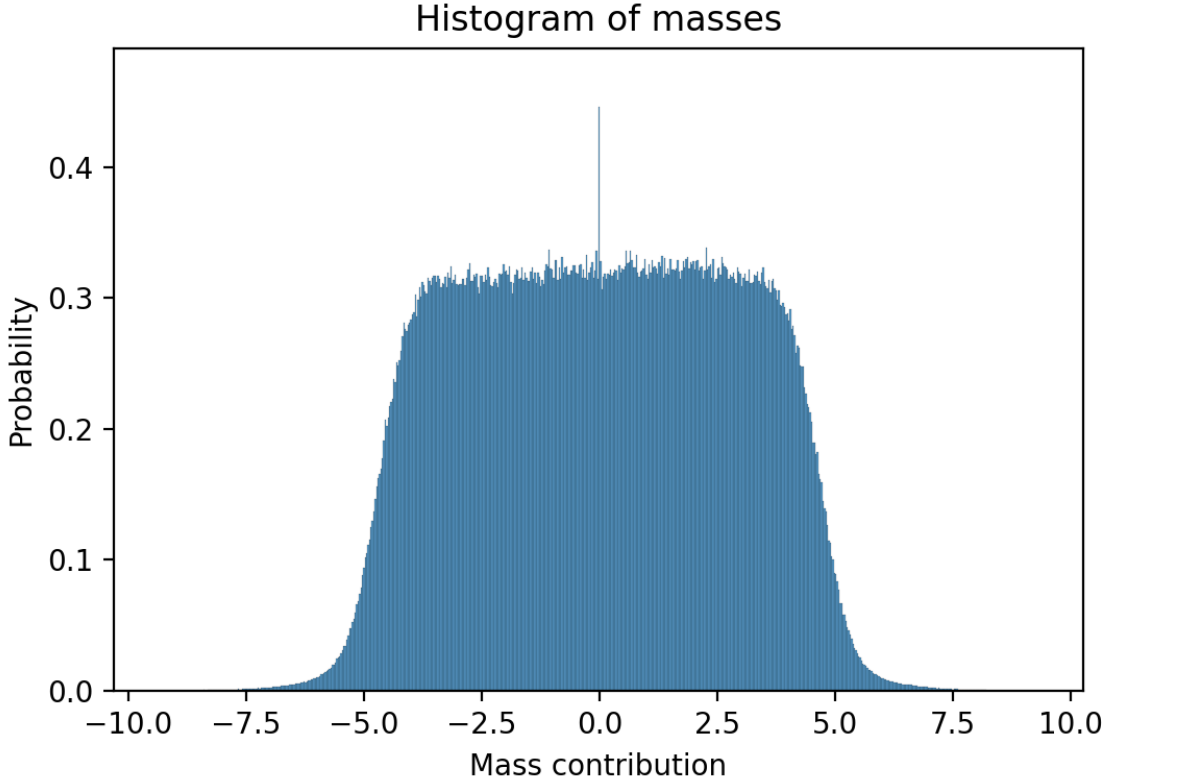}
    \caption{Histograms of the mass accrued by the time the absorbing state is reached, for an $1001$ state system as described in the text. We used $618$ bins to build this histograms. We have left the prodigal sons peak at zero in this histogram.} 
    \label{h4}
\end{figure}

The reason for this ``fractal'' peak structure that is thus revealed is that a discrete number of states generates a finite number of possible differences and so of possible masses at each step. Due to the form of $f(k)$ the first step dominates. A main set of peaks therefore arises reflecting the first step and the fact that  the differences obtainable by the largest number of pairs of states are more likely. The following steps are progressively less relevant, but determine the ancillary structure of peaks we observe, at smaller and smaller scales as we go along the chain. Since the total mass is the sum of all these contributions, its distribution is the convolution of the distributions of each contributing step. This convolution translates into a fractal structure in the histogram, erasing peak structure as we smooth, revealing it as we sharpen the bean. This is because ``smoothing'' is also a convolution: a convolution of the true underlying distribution with a smoothing window determined by the binning process. 
None of this survives the continuous limit as demonstrated in Fig.\ref{h4}, where a 1001 state system is considered covering the same range.

\begin{figure}
    \centering
    \includegraphics[width=1.05\columnwidth, height=6.7cm]{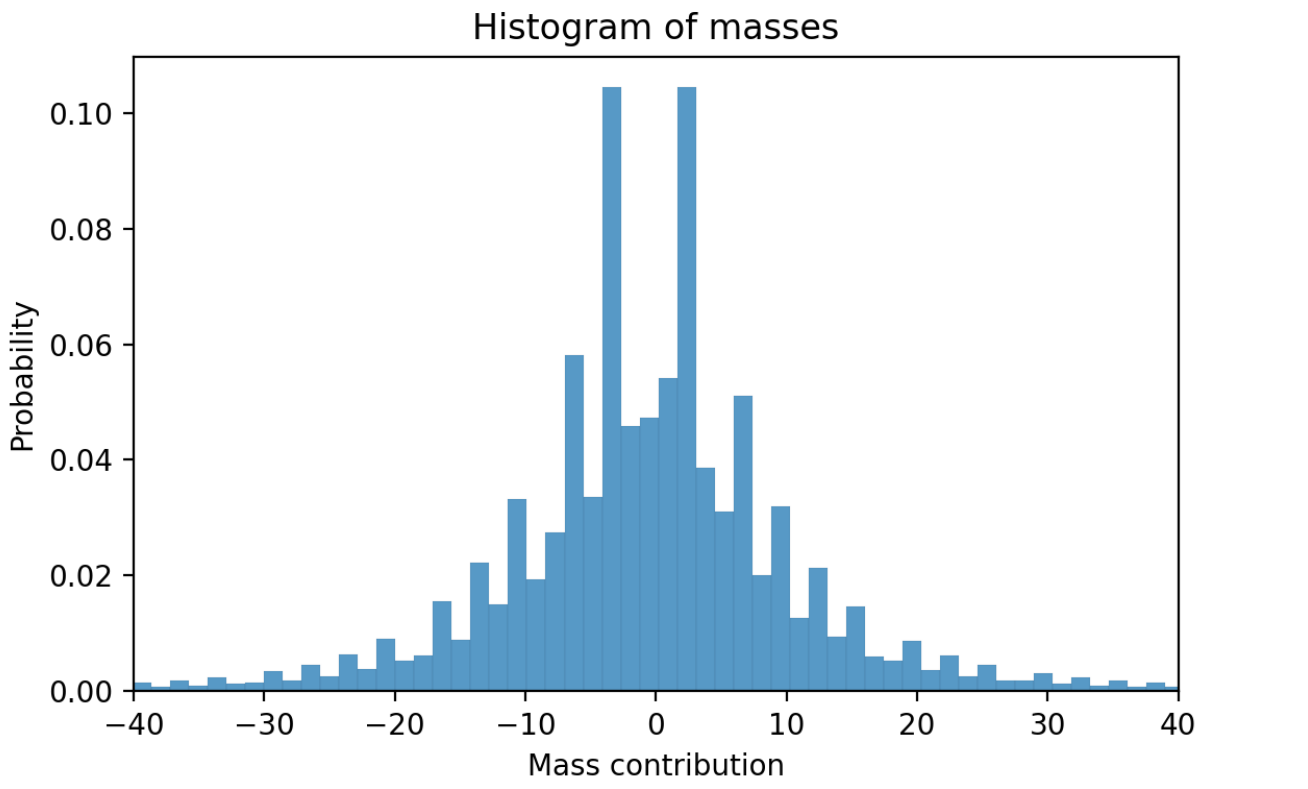}
    \caption{Probability of the final mass for a linearly increasing $f(k)$ for the same 11 state Markov matrix used before (prodigal sons have been removed).
    A leading enveloping peak roughly centered at zero is in evidence. A fractal structure of peaks can still be seen. 
    } 
    \label{HisIncf}
\end{figure}

If $f(k)$ is an increasing function the distribution is qualitatively different, as illustrated in Fig.~\ref{HisIncf}. The fractal structure of the distribution is still in evidence; however the main enveloping peak around zero is more pronounced.

\section{Conclusions}
In this paper we formalized some ideas relating evolution and chaos in the laws of physics and the emergence of our Universe~\cite{Wheeler}. Foremost, in Sections~\ref{environ}--\ref{Section:non-cons} we provided a clear deterministic definition for ``evolution'' in the laws of physics and ``time'', which generalizes~\cite{evol} beyond MSS. In our definition ``change'' in the laws of physics is akin to a chemical reaction or a non-adiabatic process, in an analogy in which the constants of nature are the counterpart of particle numbers which are conserved in adiabatic processes or non-reactive situations, and time flows at a rate defined by the chemical potential associated with these numbers. Strange as this concept might be, it simply generalizes concepts known in unimodular gravity, for the cosmological constant and unimodular time. 

The upshot is in central formula   \eqref{centralformula}, which generalizes Eq.~(82) in~\cite{evol} (valid for FRW only) to situations where homogeneity and isotropy cannot be assumed (as is presumably the case of the higgledy-piggledy phase of our Universe). As the formula indicates, matter production depends on a structural schism parting matter and gravity in terms of the origin of the cosmic clocks (i.e. the constants dual to them) and where the variability in the laws is to be found (the constants that vary with respect to these clocks). Specifically,  \eqref{centralformula} suggests a measure for separation of matter and gravity, directly related to the $NL$ (non-local or global) terms in the Poisson bracket:
\begin{equation}
    {\cal S}_{MG}\sim \{\mathbf{H}_M,\mathbf{H}_G\}_{NL}.
\end{equation}
This is the main deterministic condition for evolution to lead to the creation of the matter in our Universe. 

In Sections~\ref{higgpig1}-\ref{MassFunction} we then overlaid on this deterministic structure a random model, mimicking the random mutations powering natural selection. The central matrix $M^{IJ}\neq 0$ define in Section~\ref{higgpig1} is directly determined by ${\cal S}_{MG}$ and represents the deterministic input. The other elements determining mass creation are modeled as an absorbing Markov chain (Section~\ref{higgpig2}), where the absorbing state represent constant immutable laws. As explained in Section~\ref{higgpig1} this is required in many cases to ensure mass gains are kept. The mass outcomes are finally studied in Section~\ref{MassFunction}. 

So, in view of all of this: how to make a Universe? Here's a prescription that complies with some myths and philosophy, as well as the speculations of some respectable physicists, as discussed in the Introduction. Three stages could be distinguished. At each of these stages randomness plays a different role, as discussed in more detail in Section~\ref{higgpig1}. We may summarize them as follows. 

\begin{itemize}
    \item Stage 1: Complete chaos, undefined and incomprehensible. By the nature of the beast, there is no need to ``model'' anything about this stage or explain how it evolved to the next stage.
    \item Stage 2: The first structures appear, namely space-time, a foliation, a metric structure and conceptual separation between matter and gravity. They are described in Secs~\ref{environ}-\ref{Section:non-cons}.    Hence concepts such as diffeomorphism invariance, time (in the unimodular-like sense), deterministic time (in)-dependence of the laws, start to make sense; but are not yet realized. Instead, random mutations in the evolution  potentials and in the Hamiltonian dominate, as described in Secs.~\ref{higgpig1}-\ref{MassFunction}. 
    \item Stage 3:  Some of these processes are fortunate enough to evolve to a stage where random evolution stops, deterministic evolution is heavily suppressed (constant $\bm\beta$ potentials), and the underlying Hamiltonian is nearly diffeomorphism invariant. They correspond to the absorbing state of the Markov higgledy-piggledy chain. They also have matter, created along the chain, to be kept by the Universe from then on. 
\end{itemize}
Therefore, latter day space-time diffeomorphism symmetry (exact or very  nearly realized) is a prediction of this model of cosmogenesis. 
True, the model is predicated on breaking this symmetry, but then it only works if the symmetry exists within the set of all evolution potentials and represents the absorbing state of the model. This depends on a technicality: whether the function $f(k)$ is an increasing or a decreasing function, but just as we cannot assume a bias towards positive mass creation, we cannot assume the most favorable case for the function $f(k)$. 
Thus, one should turn off the tap of creation related to evolution.   In a state of permanent chaos, i.e. permanent random evolution, what generates positive energy could also take it away, or generate negative energy. Without a sign bias there would be an equilibrium at no matter: chaos giveth with one hand and taketh away with the other. Natural selection within the higgledy-piggledy therefore implies breaking, but then realizing the most fundamental symmetry of our world.

We defer to future work further predictions of this model. Most obviously could this process of creation be a baryogenesis explanation? Would it bypass the Sakharov conditions?

\section{Acknowledgments}
We thank C. Magueijo for assistance with terminology and Agostino Ruta for precious discussions about the results.
This work was partly supported by the STFC Consolidated Grant ST/X00575/1 (JM).


\begin{thebibliography}{0}%
\makeatletter
\providecommand \@ifxundefined [1]{%
 \@ifx{#1\undefined}
}%
\providecommand \@ifnum [1]{%
 \ifnum #1\expandafter \@firstoftwo
 \else \expandafter \@secondoftwo
 \fi
}%
\providecommand \@ifx [1]{%
 \ifx #1\expandafter \@firstoftwo
 \else \expandafter \@secondoftwo
 \fi
}%
\providecommand \natexlab [1]{#1}%
\providecommand \enquote  [1]{``#1''}%
\providecommand \bibnamefont  [1]{#1}%
\providecommand \bibfnamefont [1]{#1}%
\providecommand \citenamefont [1]{#1}%
\providecommand \href@noop [0]{\@secondoftwo}%
\providecommand \href [0]{\begingroup \@sanitize@url \@href}%
\providecommand \@href[1]{\@@startlink{#1}\@@href}%
\providecommand \@@href[1]{\endgroup#1\@@endlink}%
\providecommand \@sanitize@url [0]{\catcode `\\12\catcode `\$12\catcode `\&12\catcode `\#12\catcode `\^12\catcode `\_12\catcode `\%12\relax}%
\providecommand \@@startlink[1]{}%
\providecommand \@@endlink[0]{}%
\providecommand \url  [0]{\begingroup\@sanitize@url \@url }%
\providecommand \@url [1]{\endgroup\@href {#1}{\urlprefix }}%
\providecommand \urlprefix  [0]{URL }%
\providecommand \Eprint [0]{\href }%
\providecommand \doibase [0]{http://dx.doi.org/}%
\providecommand \selectlanguage [0]{\@gobble}%
\providecommand \bibinfo  [0]{\@secondoftwo}%
\providecommand \bibfield  [0]{\@secondoftwo}%
\providecommand \translation [1]{[#1]}%
\providecommand \BibitemOpen [0]{}%
\providecommand \bibitemStop [0]{}%
\providecommand \bibitemNoStop [0]{.\EOS\space}%
\providecommand \EOS [0]{\spacefactor3000\relax}%
\providecommand \BibitemShut  [1]{\csname bibitem#1\endcsname}%
\let\auto@bib@innerbib\@empty
\end{thebibliography}%


\begin{thebibliography}{99}

\bibitem{Hesiod}Hesiod, Theogony (e.g. 
N. Athanassakis, `` Hesiod; introduction, translation, and notes'', Baltimore: John Hopkins University Press, 1983. ISBN 0-8018-2998-4).
\bibitem{Wheeler}
John Wheeler, 
in 'The Computer and the Universe', International Journal of Theoretical Physics (1982), 21, 565. 
\bibitem{PaulDavis}
Paul Davis, https://metanexus.net/large-cosmic-lessons-physics/

\bibitem{evol}
J.~Magueijo,
Phys. Rev. D \textbf{108}, no.10, 103514 (2023)
doi:10.1103/PhysRevD.108.103514
[arXiv:2306.08390 [hep-th]].


\bibitem{unimod} M.~Henneaux and C.~Teitelboim, ``The cosmological constant and general covariance,'' {\em Phys.\ Lett.\ B} {\bf 222} (1989), 195--199.

\bibitem{unimod1}
W. G. Unruh, Phys. Rev. {\bf D40}, 1048 (1989);  K.~V.~Kucha\v{r}, ``Does an unspecified cosmological constant solve the problem of time in quantum gravity?,'' {\em Phys.\ Rev.\ D} {\bf 43} (1991), 3332--3344.


\bibitem{UnimodLee1}
L.~Smolin, ``Quantization of unimodular gravity and the cosmological constant problems,'' {\em Phys.\ Rev.\ D} {\bf 80} (2009), 084003, arXive: 0904.4841. 

\bibitem{alan} A. Daughton, J. Louko, and R. D. Sorkin, ``Instantons and unitarity in quantum cosmology with fixed four-volume,'' {\em Phys.\ Rev.\ D} {\bf 58}, 084008 (1998).

\bibitem{daughton} A. Daughton, J. Louko, and R. D. Sorkin, ``Initial conditions and unitarity in unimodular quantum cosmology,'' [gr-qc/9305016].

\bibitem{sorkin1} R. D. Sorkin, ``Role of time in the sum-over-histories framework for gravity,'' {\em Int J Theor Phys} {\bf 33}, 523–534 (1994). https://doi.org/10.1007/BF00670514

\bibitem{sorkin2} R. D. Sorkin, ``Forks in the road, on the way to quantum gravity,'' {\em Int J Theor Phys} {\bf 36}, 2759–2781 (1997). https://doi.org/10.1007/BF02435709


\bibitem{Bombelli}
L.~Bombelli, W.~E.~Couch and R.~J.~Torrence,
Phys. Rev. D \textbf{44}, 2589-2592 (1991)
doi:10.1103/PhysRevD.44.2589

\bibitem{UnimodLee2}
L.~Smolin,
Phys. Rev. D \textbf{84}, 044047 (2011)
doi:10.1103/PhysRevD.84.044047
[arXiv:1008.1759 [hep-th]].

\bibitem{JoaoLetter}
J.~Magueijo,
Phys. Lett. B \textbf{820}, 136487 (2021)
doi:10.1016/j.physletb.2021.136487
[arXiv:2104.11529 [gr-qc]].

\bibitem{JoaoPaper}
J.~Magueijo,
Phys. Rev. D \textbf{106}, no.8, 084021 (2022)
doi:10.1103/PhysRevD.106.084021
[arXiv:2110.05920 [gr-qc]].


\bibitem{pad}
N.~Kaloper and A.~Padilla,
Phys. Rev. Lett. \textbf{112}, no.9, 091304 (2014).

\bibitem{pad1}
N.~Kaloper, A.~Padilla, D.~Stefanyszyn and G.~Zahariade,
Phys. Rev. Lett. \textbf{116}, no.5, 051302 (2016)
doi:10.1103/PhysRevLett.116.051302
[arXiv:1505.01492 [hep-th]].

\bibitem{lomb}
L.~Lombriser,
Phys. Lett. B \textbf{797}, 134804 (2019).


\bibitem{vikman}
P.~Jirou\v{s}ek, K.~Shimada, A.~Vikman and M.~Yamaguchi,
JCAP \textbf{04}, 028 (2021).

\bibitem{vikman1}
A.~Vikman,
[arXiv:2107.09601 [gr-qc]].

\bibitem{vikaxion}
K.~Hammer, P.~Jirousek and A.~Vikman,
[arXiv:2001.03169 [gr-qc]].

\bibitem{geoCDM} 
J.~Magueijo,
Phys. Rev. D \textbf{109}, no.12, 124026 (2024)
doi:10.1103/PhysRevD.109.124026
[arXiv:2404.15809 [hep-th]].

\bibitem{nongeoCDM}
J.~Magueijo,
Phys. Rev. D \textbf{110}, no.8, 084050 (2024)
doi:10.1103/PhysRevD.110.084050
[arXiv:2406.17428 [gr-qc]].


\bibitem{BHevol}
J.~Magueijo,
Phys. Rev. D \textbf{109}, no.4, 044034 (2024)
doi:10.1103/PhysRevD.109.044034
[arXiv:2310.11929 [hep-th]].

\bibitem{MachianCDM}
J.~Magueijo,
Phys. Lett. B \textbf{858}, 139001 (2024)
doi:10.1016/j.physletb.2024.139001
[arXiv:2312.07597 [hep-th]].



\bibitem{VSL}
A.~Albrecht and J.~Magueijo,
Phys. Rev. D \textbf{59}, 043516 (1999)
doi:10.1103/PhysRevD.59.043516
[arXiv:astro-ph/9811018 [astro-ph]].


\bibitem{Isham}
C.~J.~Isham,
NATO Sci. Ser. C \textbf{409}, 157-287 (1993)
[arXiv:gr-qc/9210011 [gr-qc]].


\bibitem{viqar}
V.~Husain and T.~Pawlowski,
Class. Quant. Grav. \textbf{28}, 225014 (2011)
doi:10.1088/0264-9381/28/22/225014
[arXiv:1108.1147 [gr-qc]].


\bibitem{gielen} 
S.~Gielen and L.~Men\'endez-Pidal,
Class. Quant. Grav. \textbf{37}, no.20, 205018 (2020).

\bibitem{gielen1}
S.~Gielen and L.~Men\'endez-Pidal,
Class. Quant. Grav. \textbf{39}, no.7, 075011 (2022)
doi:10.1088/1361-6382/ac504f
[arXiv:2109.02660 [gr-qc]].

\bibitem{brown}
J.~D.~Brown,
Class. Quant. Grav. \textbf{10}, 1579-1606 (1993)
doi:10.1088/0264-9381/10/8/017
[arXiv:gr-qc/9304026 [gr-qc]].

\bibitem{brownkuchar}
J.~D.~Brown and K.~V.~Kuchar,
Phys. Rev. D \textbf{51}, 5600-5629 (1995)
doi:10.1103/PhysRevD.51.5600
[arXiv:gr-qc/9409001 [gr-qc]].

\bibitem{dirichlet}
Y.~W.~Teh,
Encyclopedia of machine learning \textbf{1063}, 280-287 (2010)
Citeseer

\bibitem{markov}
D.~W.~Stroock,
An introduction to Markov processes \textbf{230} (2013)
Springer Science \& Business Media

\bibitem{markov2.0}
P. Bassani and J. Magueijo, in preparation

\bibitem{Murphy:2003hw}
M.~T.~Murphy, J.~K.~Webb and V.~V.~Flambaum,
Mon. Not. Roy. Astron. Soc. \textbf{345}, 609 (2003)
doi:10.1046/j.1365-8711.2003.06970.x
[arXiv:astro-ph/0306483 [astro-ph]].

\bibitem{Webb}
J.~K.~Webb, C.~C.~Lee, D.~Milakovic, V.~V.~Flambum, V.~A.~Dzuba and J.~Magueijo,
[arXiv:2401.00888 [astro-ph.CO]].

\bibitem{newWebb}
J.~K.~Webb, C.~C.~Lee, D.~Milakovic, V.~V.~Flambum, V.~A.~Dzuba and J.~Magueijo,
[arXiv:2401.00888 [astro-ph.CO]].

\bibitem{HLPaolo}
P.~M.~Bassani, J.~Magueijo and S.~Mukohyama,
[arXiv:2408.03793 [gr-qc]].









\end{thebibliography}
\end{document}